\documentclass[aps,pra,twocolumn,10pt,nofootinbib,amsmath,amssymb]{revtex4-2}

\usepackage{amscd,graphicx,color,colordvi,bm,amsfonts,stmaryrd}
\usepackage{dsfont}
\usepackage{mathtools}
\usepackage{booktabs}
\usepackage{enumitem}
\usepackage{hyperref}
\hypersetup{hidelinks}

\DeclareMathOperator{\Tr}{Tr}

\newcommand{\FW}{F_W}
\newcommand{\z}{z}
\newcommand{\E}{\mathbb{E}}
\newcommand{\Pp}{\mathbb{P}}
\newcommand{\muF}{\mu_F}
\newcommand{\muR}{\mu_R^\Theta}
\newcommand{\GammaPath}{\Gamma}
\newcommand{\ThetaT}{\Theta}
\newcommand{\Qh}{\mathcal{Q}_{\hbar}}
\newcommand{\Acl}{A_{\rm cl}}
\newcommand{\Aw}{A_{\rm w}}
\newcommand{\Amag}{A_{\rm mag}}
\newcommand{\Asign}{A_{\rm sign}}
\newcommand{\Lcl}{L_{\rm cl}^{\dagger}}
\newcommand{\epsilonQ}{\epsilon_{\rm Q}}
\newcommand{\chiQ}{\chi_{\rm Q}}

\newenvironment{revtexlayoutpar}
  {\par\begingroup\sloppy\hbadness=10000\relax}
  {\par\endgroup}

\begin{document}

\title{Weighted Phase-Space Paths for Exact Wigner Dynamics}

\author{Surachate Limkumnerd}
\email[Corresponding author: ]{surachate.l@chula.ac.th}
\affiliation{%
Department of Physics, Faculty of Science, Chulalongkorn University, Phayathai Rd., Patumwan, Bangkok 10330, Thailand 
}
\author{Panat Phanthaphanitkul}
\email{phanthaphanitkul.panat.25x@st.kyoto-u.ac.jp}
\affiliation{%
Department of Physics No.~1, Graduate School of Science, Kyoto University, Kyoto 606-8502, Japan %
}

\date{\today}

\begin{abstract}
A quantum state can be written in phase space, but the resulting object is not generally the probability density of a positive stochastic process on ordinary phase space. We spell this out for Wigner dynamics. If a positive phase-space process is required only to reproduce the Born density after integrating over momentum, the requirement fixes only an integrated current; the local drift and diffusion remain underdetermined. If one instead requires all Weyl-ordered expectation values, the phase-space object is fixed to be the Wigner function. For non-quadratic potentials the Wigner--Moyal generator contains higher-order, signed momentum-transfer terms, so it is not the Fokker--Planck generator of a positive Brownian diffusion. The exact Wigner function must therefore be reconstructed, in a stochastic representation, as a weighted empirical measure
\[
\FW(\z,t)=\E_{\Pp}[W_t\delta(\z-\z_t)],
\qquad \z=(q,p),
\]
rather than the unweighted density of sampled carrier trajectories. With classical Hamiltonian flow as the carrier, all nonclassical correction beyond classical transport sits in the Moyal residual and can be represented by signed weights or branching events. The same split defines a residual diagnostic that vanishes for quadratic Hamiltonians and measures what classical carrier transport misses in anharmonic dynamics. The formulation also gives a forward--reverse relation for signed Wigner path measures. The ratio of forward and reversed contributions separates into a positive magnitude factor and a sign factor. This sign records the parity of the Wigner interference contribution; it is not a thermodynamic entropy production.
\end{abstract}


\maketitle

\section{Introduction}
In classical statistical mechanics, a trajectory and an ensemble density have the same probabilistic status. A realization is a path in phase space, and many such realizations define a positive phase-space density. This picture suggests a natural question: can quantum mechanics be obtained from coupled stochastic variables $(q_t,p_t)$ whose ensemble density is an ordinary probability distribution on phase space? This is the positive Langevin route to quantum phase space.

The answer is no in a precise sense. The difficulty is not that quantum mechanics lacks a phase-space description, or that stochastic constructions are unavailable. The difficulty is that the exact quantum phase-space object is not generally an ordinary probability density. If one asks only that the position marginal of a positive phase-space process be the Born density, the drift and diffusion fields are not fixed. Infinitely many positive phase-space processes can have the same configuration-space marginal. If one asks instead for the object that reproduces all Weyl-ordered expectations, the answer is fixed: it is the Wigner function. The Wigner function is equivalent to the density matrix, but it can be signed. Its evolution is the Wigner--Moyal equation, and for non-quadratic potentials that equation is not a positive second-order Fokker--Planck equation.

The point may be summarized by the incompatibility, in general, of the
three requirements
\[
\text{positivity}
+\text{phase space}
+\text{exact quantum dynamics}.
\]
Keeping positivity and exact Born statistics leads naturally to a
configuration-space process, as in Nelson stochastic mechanics
\cite{Nelson1966SchrodingerFromNewtonian,Wallstrom1989StochasticSchrodingerDerivation,Pavon1995HamiltonPrincipleStochasticMechanics}.
Keeping phase space and ordinary positivity gives either classical
transport or approximate Langevin closures.  Keeping exact quantum
dynamics together with all Weyl-ordered expectations gives the Wigner representation, and hence a
weighted or signed object.  Thus one obtains the schematic alternatives
\[
\begin{array}{@{}c@{\quad}c@{\quad}c@{}}
\begin{array}{c}
\text{positive}\\
\text{configuration-space}\\
\text{diffusion}
\end{array}
&\Longrightarrow&
\begin{array}{c}
\text{Nelson-type}\\
\text{dynamics}
\end{array}
\\[1.5em]
\begin{array}{c}
\text{positive ordinary}\\
\text{phase-space}\\
\text{process}
\end{array}
&\Longrightarrow&
\begin{array}{c}
\text{non-unique}\\
\text{approximate, or}\\
\text{special-case closure}
\end{array}
\\[1.5em]
\begin{array}{c}
\text{weighted Wigner}\\
\text{phase-space}\\
\text{measure}
\end{array}
&\Longrightarrow&
\begin{array}{c}
\text{exact}\\
\text{Wigner--Moyal}\\
\text{dynamics}
\end{array}
\end{array}
\]
The conclusion is that exact Wigner phase space can be sampled consistently only after separating the positive process being sampled from the signed phase-space object being reconstructed. The replacement used here is
\[
\FW(\z,t)=
\E_{\Pp}\!\left[W_t\delta(\z-\z_t)\right],
\qquad \z=(q,p),
\]
rather than the unweighted probability density $p_{\z_t}$ of the carrier process,
\[
\FW(\z,t)=p_{\z_t}(\z,t).
\]
The path $\z_t$ is a carrier path sampled from a positive measure $\Pp$. The quantum phase-space state is not the empirical density of these paths. It is the weighted empirical measure obtained after the weights $W_t$ are included. This distinction keeps the phase-space intuition of a stochastic construction without requiring the Wigner function itself to be positive.

The paper is organized as follows. We first show why matching the Born marginal does not determine a positive phase-space Langevin process. We then recall the Wigner transform and the equivalence between Schr\"odinger, von Neumann, and Wigner--Moyal dynamics. Comparing the Wigner--Moyal generator with a Fokker--Planck generator shows why an exact Wigner density cannot generally be an ordinary positive Brownian density on phase space. We then introduce weighted carrier paths. Choosing the carrier to be classical Hamiltonian flow splits the exact Wigner generator into the classical Liouville part and the nonclassical Moyal residual. The residual can be written as a signed kernel and represented by signed weights or branching events. The same split defines a residual diagnostic for the part of the generator omitted by classical carrier transport.

Finally, we compare forward and reversed weighted paths. Classical fluctuation relations compare positive path measures. Here the corresponding measures are signed:
\[
d\mu_F=W_F\,d\Pp_F,
\qquad
d\mu_R^\Theta=W_R^\Theta\,d\Pp_R^\Theta.
\]
On their common support,
\[
\frac{d\mu_F}{d\mu_R^\Theta}
=
\Asign e^{\Amag}.
\]
The magnitude $\Amag$ contains the ordinary carrier-law ratio and the ratio of absolute Wigner weights. The factor $\Asign$ is the sign part. It records whether a forward path and its reversed counterpart carry the same or opposite Wigner sign. We call this sign the interference parity. It is not a thermodynamic entropy production.

The ingredients are standard: the Wigner function, the Moyal bracket, and signed-particle representations of Wigner evolution all have substantial literatures \cite{Wigner1932QuantumCorrection,Moyal1949QuantumMechanicsStatisticalTheory,Hillery1984DistributionFunctionsFundamentals,SellierNedjalkovDimov2015AppliedWignerMonteCarlo,Polkovnikov2010PhaseSpaceRepresentation}. The purpose here is to put them together in a specific way: to identify where a positive Langevin construction fails, and to state the corresponding forward--reverse relation for signed Wigner paths.

\section{Why matching the Born density is not enough}

We begin with the route suggested by classical stochastic mechanics. Let $\z=(q,p)$ and consider a positive It\^o diffusion on phase space,
\[
d\z_t=A(\z_t,t)\,dt+B(\z_t,t)\,dW_t,
\]
where $A$ is a drift field, $B$ is a noise matrix, and $W_t$ is a standard Brownian motion. If $P(\z,t)$ is the probability density of $\z_t$, then $P$ obeys the Fokker--Planck equation
\[
\partial_tP
=
-\partial_i(A_iP)
+
\frac12\partial_i\partial_j(D_{ij}P),
\qquad
D=BB^T\ge0.
\]
Such an equation has two properties that will be important later: it preserves positivity and its differential order is at most two.

\begin{revtexlayoutpar}
A natural requirement is to choose $A$ and $D$ so that the position marginal agrees with the Born density,
\end{revtexlayoutpar}
\[
\rho(q,t)=|\psi(q,t)|^2,
\qquad
\rho(q,t)=\int dp\,P(q,p,t).
\]
Integrating the Fokker--Planck equation over $p$ gives
\[
\partial_t\rho(q,t)
=
-\partial_q J_q(q,t),
\]
where
\[
\begin{aligned}
J_q(q,t)
&=
\int dp\,
\bigg[
A_q(q,p,t)P(q,p,t)
\\
&\qquad
-\frac12\partial_j\!\left(D_{qj}(q,p,t)P(q,p,t)\right)
\bigg],
\end{aligned}
\]
assuming boundary terms vanish. Thus the Born continuity equation constrains only the integrated current $J_q(q,t)$, not the full phase-space current. Many different phase-space drift and diffusion fields produce the same integrated current.

This is why derivations based only on marginal matching naturally contain free functions. One can modify the phase-space current by a term whose $p$-integral vanishes, or by a divergence-free current under suitable boundary conditions, without changing the position marginal. Adding finitely many moment constraints reduces this freedom but does not remove it. The unknowns are functions on phase space, while a finite set of moment equations gives only a finite set of integral constraints.

Thus matching the Born position density, even together with finitely many phase-space moments, does not uniquely determine a positive phase-space Langevin process. The position marginal fixes only the integrated phase-space current. The full Fokker--Planck current still contains functional freedom that is invisible to a finite collection of marginal and moment equations. Current components whose contribution integrates to zero in the marginal equation may be changed without changing $\rho(q,t)$; finite moment constraints add only further integral conditions. They cannot determine the full drift field $A(q,p,t)$ and diffusion matrix $D(q,p,t)$. A constructive null-current version of this statement is given in Appendix~\ref{app:underdetermination}.

\paragraph*{Statement 1: the Born marginal does not fix the process.}
Let a positive phase-space diffusion reproduce the Born density only through its configuration-space marginal. Then the Born continuity equation fixes the $p$-integrated phase-space current, not the local drift field and diffusion matrix. Consequently no finite set of marginal or moment constraints can select a unique positive phase-space Langevin process without additional closure assumptions. This is the first limitation: before one asks for exact Wigner dynamics, the marginal problem is already underdetermined.

This is not only a technical nuisance. Agreement with $\rho(q,t)$ alone allows many positive phase-space processes. An exact quantum phase-space theory must specify a stronger target. The next section replaces marginal fitting by the Weyl--Wigner phase-space object of the quantum state.

\section{The Wigner function as the phase-space target}
\label{sec:wigner_representation}

A quantum phase-space object should not be defined by fitting a few marginals. It should reproduce the quantum state in phase space. The standard object with this property is the Wigner function \cite{Wigner1932QuantumCorrection,Hillery1984DistributionFunctionsFundamentals}.

For a density operator $\hat\rho(t)$, define
\[
\FW(q,p,t)
=
\frac{1}{2\pi\hbar}
\int dy\,e^{-ipy/\hbar}
\left\langle q+\frac y2\middle|\hat\rho(t)\middle|q-\frac y2\right\rangle.
\]
For a pure state $\psi(q,t)$, this becomes
\[
\begin{aligned}
\FW(q,p,t)
&=
\frac{1}{2\pi\hbar}
\int dy\,e^{-ipy/\hbar}
\\
&\qquad{}\times
\psi^*\!\left(q-\frac y2,t\right)
\psi\!\left(q+\frac y2,t\right).
\end{aligned}
\]

The Wigner function has the correct position and momentum marginals:
\[
\int dp\,\FW(q,p,t)=|\psi(q,t)|^2,
\]
and
\[
\int dq\,\FW(q,p,t)=|\tilde\psi(p,t)|^2.
\]
More importantly, it reproduces Weyl-ordered expectation values. If $A_W(q,p)$ is the Weyl symbol of an operator $\hat A$, then
\[
\langle \hat A\rangle
=
\int dq\,dp\,\FW(q,p,t)A_W(q,p).
\]
This condition is stronger than matching a finite set of moments: it fixes the phase-space object associated with the Weyl correspondence.

The Wigner transform is also invertible. One has
\[
\left\langle x\middle|\hat\rho(t)\middle|x'\right\rangle
=
\int dp\,
e^{ip(x-x')/\hbar}
\FW\!\left(\frac{x+x'}2,p,t\right).
\]
Thus $\FW$ contains exactly the same information as $\hat\rho$. In a pure state, this is the same physical information as $\psi$, up to a global phase.

The dynamical equivalence is equally direct. The von Neumann equation
\[
i\hbar\partial_t\hat\rho=[\hat H,\hat\rho]
\]
is mapped by the Weyl transform into the Wigner--Moyal equation
\[
\partial_t\FW=\{H,\FW\}_M.
\]
For pure states, the chain of equivalences is
\[
\begin{aligned}
i\hbar\partial_t\psi=\hat H\psi
&\Longleftrightarrow
i\hbar\partial_t\hat\rho=[\hat H,\hat\rho]
\\
&\Longleftrightarrow
\partial_t\FW=\{H,\FW\}_M.
\end{aligned}
\]

Thus the Wigner--Moyal equation is the Weyl transform of the von Neumann equation, and for pure states it is equivalent to Schr\"odinger evolution up to a global phase. This equivalence fixes the target of an exact phase-space theory. Once the phase-space object is required to reproduce all Weyl-ordered expectations, $F(q,p,t)$ is no longer an arbitrary density to be engineered by a Langevin equation. It is the Wigner function, and it is generally not positive. Wigner negativity is therefore not a flaw in the representation. It is the sign that exact quantum phase space is not an ordinary probability space. Hudson's theorem makes this explicit for pure continuous-variable states: nonnegative Wigner functions are Gaussian. Related finite-dimensional and mixed-state extensions further show how restrictive Wigner nonnegativity is \cite{Hudson1974WignerNonnegative,Gross2006FiniteHudson,Mandilara2009ExtendingHudson}.

\section{Why a positive Fokker--Planck equation fails}

We now compare the exact Wigner generator with the generator of a positive Langevin process. Let
\[
H(q,p)=\frac{p^2}{2m}+V(q).
\]
The Wigner--Moyal equation gives
\[
\partial_t\FW
=
-\frac{p}{m}\partial_q\FW
+
V'(q)\partial_p\FW
+
\Qh[\FW],
\]
where the nonclassical residual is
\[
\Qh[\FW]
=
\sum_{n=1}^{\infty}
\frac{(-1)^n}{(2n+1)!}
\left(\frac{\hbar}{2}\right)^{2n}
V^{(2n+1)}(q)\partial_p^{2n+1}\FW.
\]
The first two terms form the classical Liouville generator. The remaining terms are the Moyal corrections.

For a quadratic potential, all derivatives $V^{(k)}$ with $k\ge3$ vanish, and hence
\[
\Qh=0.
\]
This is why harmonic oscillator Wigner dynamics is exactly classical phase-space rotation, even for nonclassical states. The nonclassicality of such states is carried by the initial Wigner function itself, not by a quantum correction to the phase-space flow.

For non-quadratic potentials, however, $\Qh$ contains higher odd derivatives in momentum. The leading correction is
\[
-\frac{1}{3!}
\left(\frac{\hbar}{2}\right)^2
V'''(q)\partial_p^3\FW.
\]
This term is already incompatible with the generator of an ordinary Brownian diffusion, whose Fokker--Planck operator has at most second derivatives and whose diffusion matrix is positive semidefinite.

For generic non-quadratic potentials, the Wigner--Moyal generator therefore cannot be the generator of an ordinary positive Brownian diffusion on phase space with probability density $\FW$. An ordinary It\^o diffusion has a Fokker--Planck generator
\[
L_{\rm FP}^\dagger P
=
-\partial_i(A_iP)
+
\frac12\partial_i\partial_j(D_{ij}P),
\qquad
D\ge0,
\]
which is a second-order positivity-preserving operator. The Wigner--Moyal generator for a non-quadratic potential contains higher odd derivatives, or equivalently signed nonlocal momentum-transfer kernels. The Wigner function evolved by this generator is also generally signed. Thus $\FW$ cannot generally be the probability density of an ordinary positive Brownian phase-space diffusion. Special quadratic cases, approximate closures, and enlarged representations are distinguished in Appendix~\ref{app:fokker_planck_limit}.

\paragraph*{Statement 2: a positive Fokker--Planck density is not exact Wigner dynamics.}
For a Hamiltonian $H=p^2/2m+V(q)$ with a nonzero higher odd derivative of $V$, the Wigner--Moyal generator is not the adjoint generator of an ordinary positive Brownian diffusion whose density is $\FW$. The reason is local at the level of generators: a Brownian Fokker--Planck operator is second order and positivity preserving, whereas the Wigner--Moyal operator contains higher odd momentum derivatives, or equivalently signed momentum-transfer kernels. This statement leaves intact signed-particle, weighted, complex, enlarged, and approximate phase-space constructions; it excludes only the unweighted positive Brownian-density interpretation of the Wigner function.

This is the sense in which the positive phase-space Langevin program fails as an exact Wigner-density interpretation. It does not rule out all stochastic phase-space representations. It rules out interpreting the exact Wigner function as an ordinary positive Brownian trajectory density. The appropriate replacement is a weighted empirical measure, introduced next.

\section{Weighted paths for Wigner dynamics}

The limitation above does not say that stochastic phase-space representations are impossible. It says that the exact Wigner function cannot generally be the unweighted positive density of an ordinary Brownian phase-space process. The natural replacement is to separate the path being sampled from the signed object being reconstructed.

Let $\Omega$ be a path space whose elements $\GammaPath$ contain a phase-space carrier trajectory $\z_t=(q_t,p_t)$, and, when needed, additional marks such as jumps or branching events. Let $\Pp$ be a positive probability measure on $\Omega$. A weighted Wigner path representation is a pair $(\Pp,W)$ such that, for suitable test functions $\phi$,
\[
\int d\z\,\FW(\z,T)\phi(\z)
=
\E_{\Pp}
\left[
W[\GammaPath]\phi(\z_T)
\right].
\]
Equivalently, in a distributional notation,
\[
\FW(\z,T)
=
\E_{\Pp}
\left[
W[\GammaPath]\delta(\z-\z_T)
\right].
\]
The weight $W[\GammaPath]$ may be signed, and in more general representations may be complex. Here we focus on real signed Wigner weights.

\paragraph*{Statement 3: sample positively, reconstruct with signs.}
At a fixed time, if the signed Wigner measure $\FW(z,T)\,dz$ is dominated by a positive terminal sampling law $p_T(z)\,dz$, then
\[
W_T(z)=\frac{\FW(z,T)}{p_T(z)}
\]
gives $\FW(z,T)=\E_{p_T}[W_T\delta(z-z_T)]$ in the sense of distributions. In a path representation this terminal identity is implemented dynamically by the carrier law and the residual weight or branching rule. The initial Wigner sign, when present, must likewise be represented by an initial weight or positive envelope; it cannot be treated as an initial positive probability density.

This is the sense in which a stochastic phase-space representation remains possible. The carrier histories are sampled from an ordinary positive probability measure $\Pp$, but the quantum state is not their empirical density. It is their weighted empirical measure. The carrier paths are therefore not trajectories whose unweighted distribution is the quantum phase-space state. They are a way to reconstruct a quasiprobability.

Let $L_W^\dagger$ denote the Wigner--Moyal generator and let $L_0^\dagger$ denote the generator of the chosen carrier process. Formally,
\[
L_W^\dagger
=
L_0^\dagger+\mathcal R.
\]
The residual $\mathcal R$ is represented by the evolution of weights, by signed jumps, or by branching events. This is the general form of the replacement:
\[
\begin{array}{@{}c@{\quad}c@{\quad}c@{}}
\begin{array}{c}
\text{positive carrier paths}\\[-1pt]
+\ \text{signed weight}
\end{array}
&
\Longrightarrow
&
\text{exact Wigner dynamics}.
\end{array}
\]

There is still a choice in $L_0^\dagger$. If it were chosen arbitrarily, the framework would be too broad. The next section fixes a natural carrier by requiring a transparent classical limit.

\section{Classical flow and the missing quantum term}

We take the carrier process to be the classical Hamiltonian flow generated by the same Hamiltonian function,
\[
H(q,p)=\frac{p^2}{2m}+V(q).
\]
Thus
\[
\dot q=\frac{p}{m},
\qquad
\dot p=-V'(q).
\]
The associated Liouville generator is
\[
L_{\rm cl}^\dagger F
=
-\frac{p}{m}\partial_qF+V'(q)\partial_pF.
\]
The Wigner generator then splits as
\[
L_W^\dagger
=
L_{\rm cl}^\dagger+\Qh,
\]
where
\[
\Qh F
=
\sum_{n=1}^{\infty}
\frac{(-1)^n}{(2n+1)!}
\left(\frac{\hbar}{2}\right)^{2n}
V^{(2n+1)}(q)\partial_p^{2n+1}F.
\]

This split is useful for three reasons. First, it uses the same Hamiltonian that defines the quantum dynamics. Second, it makes the classical limit explicit:
\[
\Qh\to0
\qquad
\text{as}
\qquad
\hbar\to0,
\]
provided the Wigner function approaches a regular classical phase-space density. Third, it puts the nonclassical dynamical correction in the residual term $\Qh$. The carrier motion is classical; the quantum correction is carried by signed weights or branching.

With classical Hamiltonian flow as carrier, the residual generator is exactly the nonclassical Moyal correction. In the formal classical limit $\hbar\to0$, this residual vanishes and the weighted representation reduces to classical Liouville transport whenever the Wigner function approaches a positive classical density.

This does not assert that the classical carrier is the only possible sampling process. Other carriers may be useful for variance reduction or numerical efficiency. The claim is that, for an exact phase-space representation with a transparent classical limit, classical Hamiltonian flow is the natural reference relative to which the quantum residual is measured.

\section{Signed source and sink activity}

The residual $\Qh$ can be represented in differential form, as above, or in an equivalent signed-kernel form. Schematically, write
\[
\Qh F(q,p)
=
\int dp'\,K_\hbar(q,p,p')F(q,p').
\]
The precise kernel depends on the convention used to separate the classical Liouville part from the Wigner potential operator. The important point here is that the residual kernel is generally signed. It cannot be interpreted as an ordinary positive transition rate.

Once a signed residual kernel or signed measure has been fixed, its Hahn--Jordan decomposition gives two positive components,
\[
K_\hbar=K_\hbar^+-K_\hbar^-,
\qquad K_\hbar^\pm\ge0 .
\]
When this signed measure has an ordinary density $K_\hbar$, the decomposition reduces pointwise to
\[
K_\hbar^+=\max(K_\hbar,0),\qquad
K_\hbar^-=\max(-K_\hbar,0).
\]
After such a signed residual representation has been fixed, the two positive components can be used as event rates carrying opposite signs. In a signed branching representation, events generated by $K_\hbar^+$ and $K_\hbar^-$ contribute with opposite signs. The resulting ensemble reconstructs the action of the residual Wigner generator in expectation.

Signed-particle and branching-random-walk representations of the Wigner equation are established in the Wigner Monte Carlo literature \cite{Kosina2003MonteCarloCarrierTransport,Nedjalkov2004UnifiedParticleWignerBoltzmann,Sellier2015SignedParticleFormulation,ShaoXiong2020BranchingRandomWalkWigner,Muscato2021WignerEnsembleNoSplitting}. We use them here for a different purpose. They show what replaces the interpretation of the exact Wigner function as a positive Langevin phase-space density once the exact Wigner generator has been fixed.

For a fixed signed kernel or signed measure $K_\hbar$, the Hahn--Jordan decomposition minimizes total variation activity among positive decompositions of that object. In the density case, if $K_\hbar=K_1-K_2$ with $K_1,K_2\ge0$, then $K_1+K_2\ge |K_\hbar|$ pointwise, and the Hahn--Jordan choice attains equality. More generally, the total variation measure is minimal in the usual measure-theoretic sense. The measure-theoretic form of this argument is given in Appendix~\ref{app:jordan_activity}.

This minimality should not be confused with numerical optimality. The Hahn--Jordan split may lead to severe sign growth, and other sampling schemes may reduce variance. Its role here is to give a reference decomposition, not the best algorithm: for a fixed residual kernel, it is the least total signed activity needed to write that kernel as the difference of positive event rates. It is therefore a reference activity scale for signed-particle sampling schemes, not a universal lower bound on estimator variance.

\section{Reversing signed Wigner paths}

We now formulate the forward--reverse comparison for weighted Wigner paths. This is the signed analogue of the ordinary probability ratio used in classical fluctuation relations.

Let $\Pp_F$ be a positive sampling law for forward carrier histories $\GammaPath$, and let $W_F[\GammaPath]$ be the corresponding signed Wigner weight. Define the forward signed path measure
\[
d\muF(\GammaPath)
=
W_F[\GammaPath]\,d\Pp_F(\GammaPath).
\]
For a reversed protocol, define a positive sampling law $\Pp_R$ and a signed weight $W_R$. Let time reversal act on phase space by
\[
\ThetaT(q,p)=(q,-p),
\]
with external protocols reversed when present. The reversed path $\GammaPath^\dagger$ is the time reversal of $\GammaPath$. Pull the reversed signed measure back to the forward path space:
\[
d\muR(\GammaPath)
=
W_R[\GammaPath^\dagger]\,d\Pp_R^\Theta(\GammaPath),
\]
where $d\Pp_R^\Theta(\GammaPath)=d\Pp_R(\GammaPath^\dagger)$.

Assume that the two signed measures are comparable on a common support, and restrict attention to paths for which both weights are nonzero. This is the signed-measure analogue of the absolute-continuity assumption used in ordinary fluctuation relations. The restriction is important: paths with zero or singular reversed contribution must be treated as singular components rather than folded into the logarithmic magnitude term. Then
\[
\frac{d\muF}{d\muR}[\GammaPath]
=
\frac{W_F[\GammaPath]}{W_R[\GammaPath^\dagger]}
\frac{d\Pp_F}{d\Pp_R^\Theta}[\GammaPath].
\]
Because the weights may be negative, we do not take a logarithm of the signed ratio. Instead write
\[
W_F=\operatorname{sgn}(W_F)|W_F|,
\qquad
W_R^\Theta=\operatorname{sgn}(W_R^\Theta)|W_R^\Theta|.
\]
This gives
\[
\frac{d\muF}{d\muR}[\GammaPath]
=
\Asign[\GammaPath]\,
e^{\Amag[\GammaPath]},
\]
where
\[
\Amag[\GammaPath]
=
\log
\frac{d\Pp_F}{d\Pp_R^\Theta}[\GammaPath]
+
\log
\frac{|W_F[\GammaPath]|}{|W_R[\GammaPath^\dagger]|},
\]
and
\[
\Asign[\GammaPath]
=
\operatorname{sgn}(W_F[\GammaPath])
\operatorname{sgn}(W_R[\GammaPath^\dagger]).
\]

For mutually comparable forward and reversed signed Wigner path measures, the ratio therefore has the form
\[
\frac{d\muF}{d\muR} = \Asign e^{\Amag}.
\]
The magnitude term $\Amag$ contains the ordinary carrier-law asymmetry and the absolute weight asymmetry. The factor $\Asign$ is a sign factor, absent from ordinary positive path-probability ratios.

A corresponding integral identity follows formally. Since
\[
d\muR
=
\Asign e^{-\Amag}d\muF,
\]
one obtains
\[
\int d\muR
=
\int \Asign e^{-\Amag}\,d\muF.
\]
When the singular parts outside the common support $\Omega_0$ vanish and the two signed measures have the same total normalization, this gives the normalized signed integral
\[
\frac{\int_{\Omega_0}\Asign e^{-\Amag}\,d\muF}
{\int_{\Omega_0}d\muF}
=1 .
\]
Equivalently, using the positive carrier measure,
\[
\int_{\Omega_0}
W_F[\GammaPath]\Asign[\GammaPath]e^{-\Amag[\GammaPath]}
\,d\Pp_F
=
\int_{\Omega_0}
W_R^\Theta[\GammaPath]\,d\Pp_R^\Theta .
\]
This is the corresponding signed-measure analogue of an integral fluctuation relation. It is not, by itself, a thermodynamic entropy-production theorem.

The point of the decomposition is to separate forward--reverse Wigner path comparison into a magnitude part and a sign part. Classical path ratios involve only positive probabilities. Here the sign factor records whether forward and reversed weighted histories contribute with the same or opposite Wigner sign.

\section{Interference signs, not entropy production}

The relation in the previous section has the formal appearance of a fluctuation relation, but its interpretation is different from the usual thermodynamic one. In stochastic thermodynamics, the ratio
\[
\frac{d\Pp_F}{d\Pp_R^\Theta}
\]
compares two positive path measures. Its logarithm can often be identified with entropy production, heat dissipation, or work-like irreversibility, depending on the physical setup. Here the object being compared is not merely a positive path probability. It is a signed Wigner path contribution,
\[
d\mu_F=W_F\,d\Pp_F.
\]
The forward--reverse comparison therefore contains more structure than a classical path-probability ratio.

The magnitude term decomposes as
\[
\Amag=\Acl+\Aw,
\]
with
\[
\Acl
=
\log
\frac{d\Pp_F}{d\Pp_R^\Theta},
\qquad
\Aw
=
\log
\frac{|W_F|}{|W_R^\Theta|}.
\]
The first term is the ordinary carrier-path asymmetry. The second term is an asymmetry in the magnitude of the Wigner correction. The remaining factor,
\[
\Asign
=
\operatorname{sgn}(W_F)\operatorname{sgn}(W_R^\Theta),
\]
has no analogue in a theory of positive path probabilities. It records whether the forward and reversed path contributions carry the same or opposite sign.

We interpret this sign as an interference parity. A negative sign does not mean a negative frequency of occurrence. It means that the corresponding path contribution enters destructively in the reconstruction of the Wigner quasiprobability. Thus the signed Wigner path relation compares not only how often carrier histories occur, but also how their signed contributions transform under time reversal.

This is the physical content of the relation:
\[
\begin{array}{@{}c@{}}
\text{classical path asymmetry}
\longrightarrow
\text{carrier asymmetry}
\\[2pt]
+
\text{weight asymmetry}
+
\text{interference sign}.
\end{array}
\]
In regimes where the Wigner function approaches a positive classical density and the residual weights become trivial,
\[
\Asign\to1,
\qquad
\Aw\to0,
\]
and the relation reduces to the ordinary logic of positive path-probability ratios. Away from that limit, forward--reverse accounting in phase space involves not only positive likelihood ratios but also signed cancellation.

This distinction also separates static Wigner nonclassicality from dynamically generated Wigner correction. Static measures such as Wigner negativity volume quantify the signed structure of a state at one time \cite{KenfackZyczkowski2004NegativityIndicator}, and more general phase-space criteria can reveal nonclassicality even beyond direct Wigner negativity \cite{BohmannAgudeloSperling2020PhaseSpaceMatrices}. The sign factor introduced here is different: it belongs to a comparison of weighted path contributions. If it is used as a future diagnostic of dynamical nonclassicality, the initial Wigner sign must be separated from the sign generated by the residual dynamics. The harmonic oscillator illustrates the point. A non-Gaussian superposition may have a negative Wigner function, but because $\Qh=0$ its Wigner evolution is classical rotation; there is static negativity but no dynamically generated Moyal residual.

For this reason we do not identify $\Amag$ or $\Asign$ with thermodynamic entropy production. The systems considered here are closed Hamiltonian quantum systems unless otherwise stated. No heat bath, measurement record, or thermodynamic work protocol has been introduced. The relation is a signed-measure identity for exact Wigner phase-space dynamics. It is closer in spirit to quasiprobability fluctuation frameworks, where negative or signed contributions signal genuinely quantum features \cite{Hofer2017QuasiprobabilityDynamicObservables,Lostaglio2018QuantumFluctuationContextuality,Levy2020QuasiprobabilityHeatFluctuations,Lostaglio2023KirkwoodDiracQuasiprobability}, than to ordinary entropy-production theorems. Wigner-space entropy-production fluctuation theorems exist in thermodynamic settings \cite{Deffner2013QuantumEntropyPhaseSpace}; the present object is different because it is defined directly from the signed path representation of unitary Wigner--Moyal dynamics.

Equivalently, one may summarize the distinction as
\[
\begin{array}{@{}c@{}}
\text{classical fluctuation relations quantify}
\\
\text{probability imbalance;}
\\[2pt]
\text{the signed Wigner relation quantifies}
\\
\text{interference balance.}
\end{array}
\]
More precisely, the sign factor $\Asign$ identifies the part of the forward--reverse accounting carried by signed Wigner interference rather than by positive probability alone.

\section{Examples}

Two simple Hamiltonians clarify the role of the residual Moyal term. The harmonic oscillator is the null case: the exact Wigner dynamics is classical phase-space rotation, so no signed residual branching is generated dynamically. A weakly non-quadratic oscillator is the first case where a positive Fokker--Planck closure fails dynamically and a signed residual evolution is needed.

\subsection{Harmonic oscillator: exact rotation}

Let
\[
V(q)=\frac12m\omega^2q^2.
\]
Since all derivatives $V^{(k)}$ with $k\ge3$ vanish, the Moyal residual is zero:
\[
\Qh=0.
\]
The Wigner equation is therefore exactly the classical Liouville equation,
\[
\partial_t\FW
=
-\frac{p}{m}\partial_q\FW
+
m\omega^2q\,\partial_p\FW.
\]
The solution is rigid rotation in phase space. If $\Phi_t$ denotes the classical oscillator flow, then
\[
\FW(z,t)=\FW(\Phi_{-t}z,0).
\]

This remains true even when the initial state is nonclassical. For example, consider
\[
|\psi(t)\rangle
=
\frac{1}{\sqrt2}
\left(
|0\rangle e^{-iE_0t/\hbar}
+
|2\rangle e^{-iE_2t/\hbar}
\right).
\]
The initial Wigner function contains interference structure, and may take negative values. But its evolution is still exact phase-space rotation. The position density is recovered by marginalization:
\[
\rho(q,t)=\int dp\,\FW(q,p,t).
\]
No tunable diffusion coefficient is required.

This observation is important for interpreting positive diffusion simulations of oscillator superpositions. For a quadratic Hamiltonian, any discrepancy with the exact density comes from the chosen closure or representation, not from a missing diffusion coefficient in the exact Wigner dynamics. The exact phase-space dynamics is already known: rotate the Wigner function and marginalize. A mismatch means that the simulated object was not the exact Wigner function.

\subsection{Quartic oscillator: where the residual appears}

Now consider a weakly non-quadratic oscillator,
\[
V(q)=\frac12m\omega^2q^2+\lambda q^4.
\]
Then
\[
V'''(q)=24\lambda q.
\]
The leading Moyal correction is
\[
-\frac{1}{3!}
\left(\frac{\hbar}{2}\right)^2V'''(q)\partial_p^3\FW
=
-\hbar^2\lambda q\,\partial_p^3\FW.
\]
This term is third order in momentum derivatives. It cannot be generated by an ordinary positive Brownian phase-space diffusion. It is the simplest explicit example of the Fokker--Planck limitation above.

In the weighted-path formulation, the quartic correction belongs to the residual $\Qh$. If represented in kernel form, it corresponds to signed momentum-transfer or branching structure. This is where the weighted representation becomes dynamically essential for exact Wigner phase space, rather than merely transporting an initially signed Wigner function. The numerical benchmarks below isolate this residual against an exact Schr\"odinger reference.

\section{Numerical benchmarks}

The following deterministic grid benchmarks are not proposed as a new Wigner solver. Their purpose is narrower: they test the split developed above between classical carrier transport and the signed Moyal residual. We use dimensionless units $\hbar=m=\omega=1$ and the initial state
\[
|\psi_0\rangle=\frac{|0\rangle+|2\rangle}{\sqrt{2}}.
\]
The reference evolution is obtained from the time-dependent Schr\"odinger equation and converted to the Wigner convention
\[
W(q,p)
=
\frac{1}{2\pi}\int dy\,e^{-ipy}
\psi(q+y/2)\psi^*(q-y/2).
\]
All main benchmarks are evaluated at $t_f=\pi/2$.

\subsection{Quadratic dynamics: the null test}

For the harmonic Hamiltonian
\[
H=\frac{p^2}{2}+\frac{q^2}{2},
\]
the Moyal residual vanishes exactly. The Wigner function is therefore transported by the classical phase-space flow, even when the initial Wigner function is non-Gaussian and signed. This is the required null test: any residual activity in the quadratic case would be a numerical or representational artifact, not quantum anharmonic dynamics.

\begin{figure*}[t]
\includegraphics[width=\textwidth]{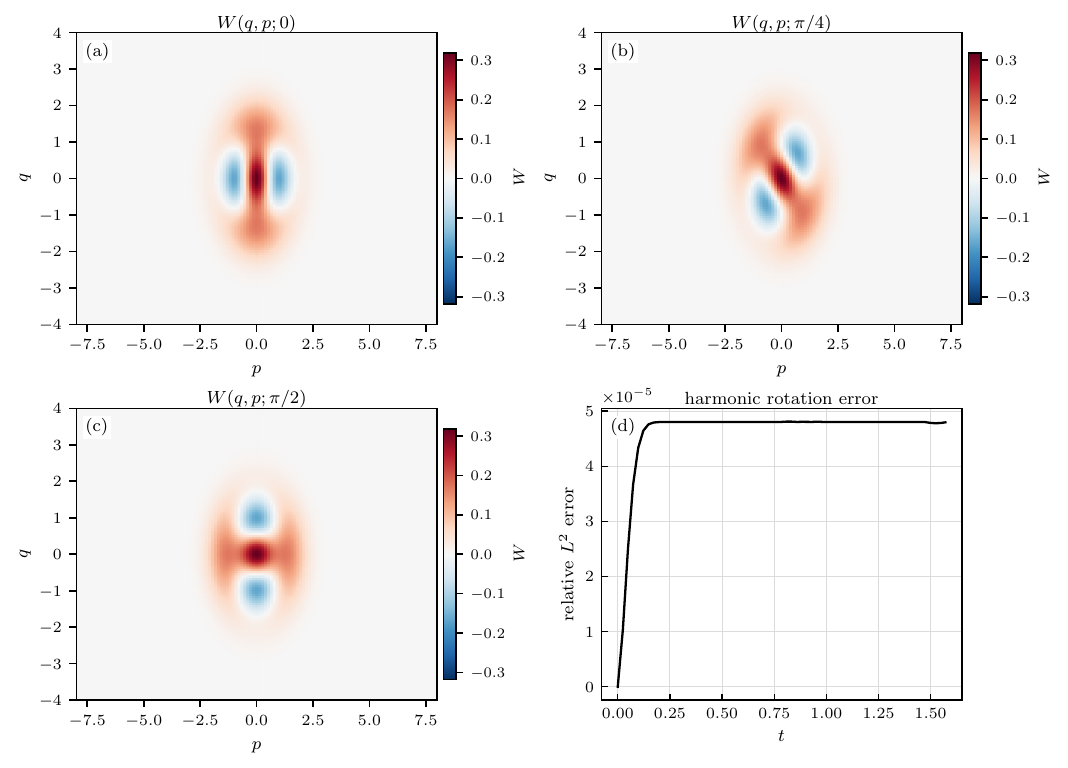}
\caption{Harmonic oscillator null test for the initial state $(|0\rangle+|2\rangle)/\sqrt{2}$. Although the initial Wigner function is non-Gaussian and signed, quadratic evolution transports it by rigid classical phase-space rotation. The Moyal residual vanishes, and the measured rotation error remains at the interpolation floor of the balanced grid.}
\label{fig:harmonic_null_test}
\end{figure*}

Figure~\ref{fig:harmonic_null_test} shows the point visually. Panels (a)--(c) are not three different approximations; they are the same signed Wigner pattern carried by the oscillator rotation. The positive and negative lobes rotate together without any dynamically generated source term. Panel (d) quantifies the remaining discrepancy between the grid-transported result and the exact rotation. Its saturation at about $4.8\times10^{-5}$ identifies the interpolation and grid floor of the balanced benchmark. Thus the benchmark confirms the null prediction $\Qh=0$: the carrier flow alone is exact for a quadratic Hamiltonian, apart from numerical interpolation error.

\subsection{Anharmonic dynamics: the missing residual}

For the quartic benchmark we use
\[
V(q)=\frac{q^2}{2}+\lambda q^4,
\qquad
\lambda=0.02.
\]
In the units used here, the exact quartic Moyal residual is
\[
R_{\rm Moyal}
=
-\lambda q\,\partial_p^3 W .
\]
Classical Liouville transport of the same initial Wigner function omits this finite signed residual. The resulting error is therefore not a failure of the Wigner convention, marginal normalization, or boundary handling; it is precisely the nonclassical part of the Wigner--Moyal generator isolated by the carrier-residual split.

\begin{figure*}[t]
\includegraphics[width=\textwidth]{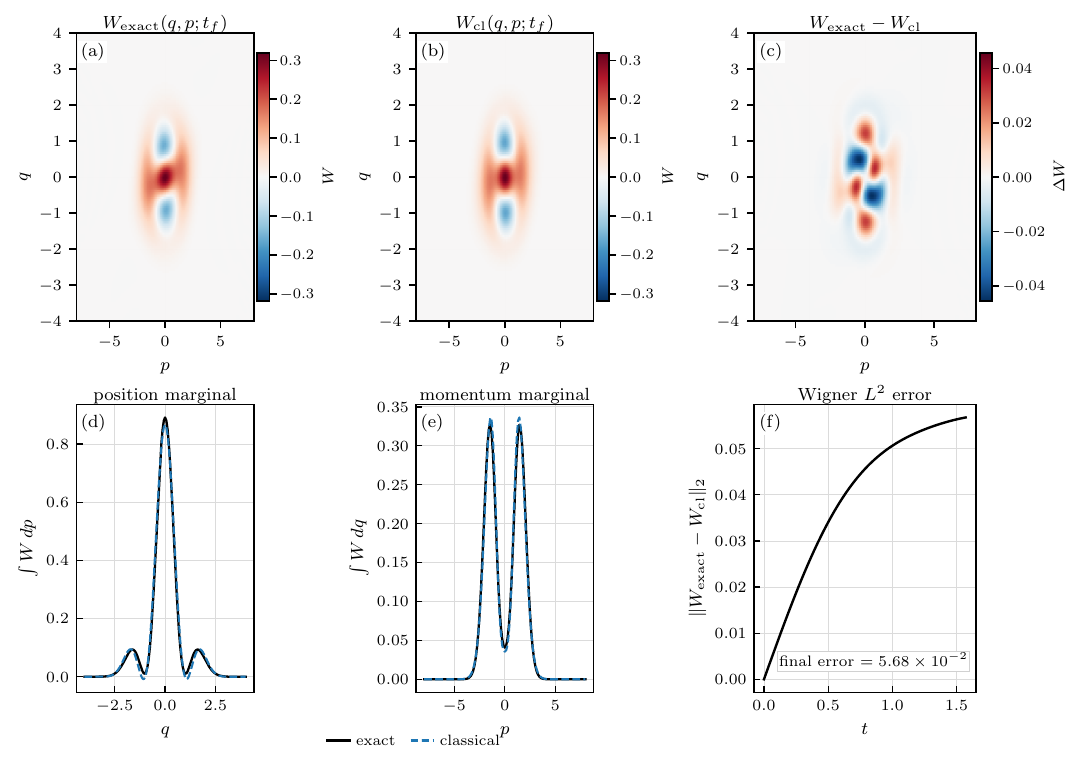}
\caption{Breakdown of classical carrier transport for $V(q)=q^2/2+\lambda q^4$ with $\lambda=0.02$. The exact Wigner evolution obtained from the Schr\"odinger reference solution is compared with classical Liouville transport of the same initial Wigner function. The discrepancy is the finite Moyal residual generated by the anharmonic term, not a boundary artifact or normalization error.}
\label{fig:quartic_exact_vs_classical}
\end{figure*}

Figure~\ref{fig:quartic_exact_vs_classical} displays the consequence of omitting this residual. Panels (a) and (b) show that the exact and classical-carrier Wigner functions remain similar at the coarse level, but panel (c) reveals a structured signed difference concentrated around the interference lobes. The position and momentum marginals in panels (d) and (e) remain close on the plotted scale, while panel (f) shows that the full Wigner $L^2$ error grows to about $5.7\times10^{-2}$ by $t_f$. This combination is important for the argument of the paper: agreement of low-dimensional marginals can coexist with a definite phase-space error. The benchmark therefore illustrates, in a concrete anharmonic evolution, why marginal matching is weaker than reproducing the full Wigner function.

\begin{figure*}[t]
\includegraphics[width=\textwidth]{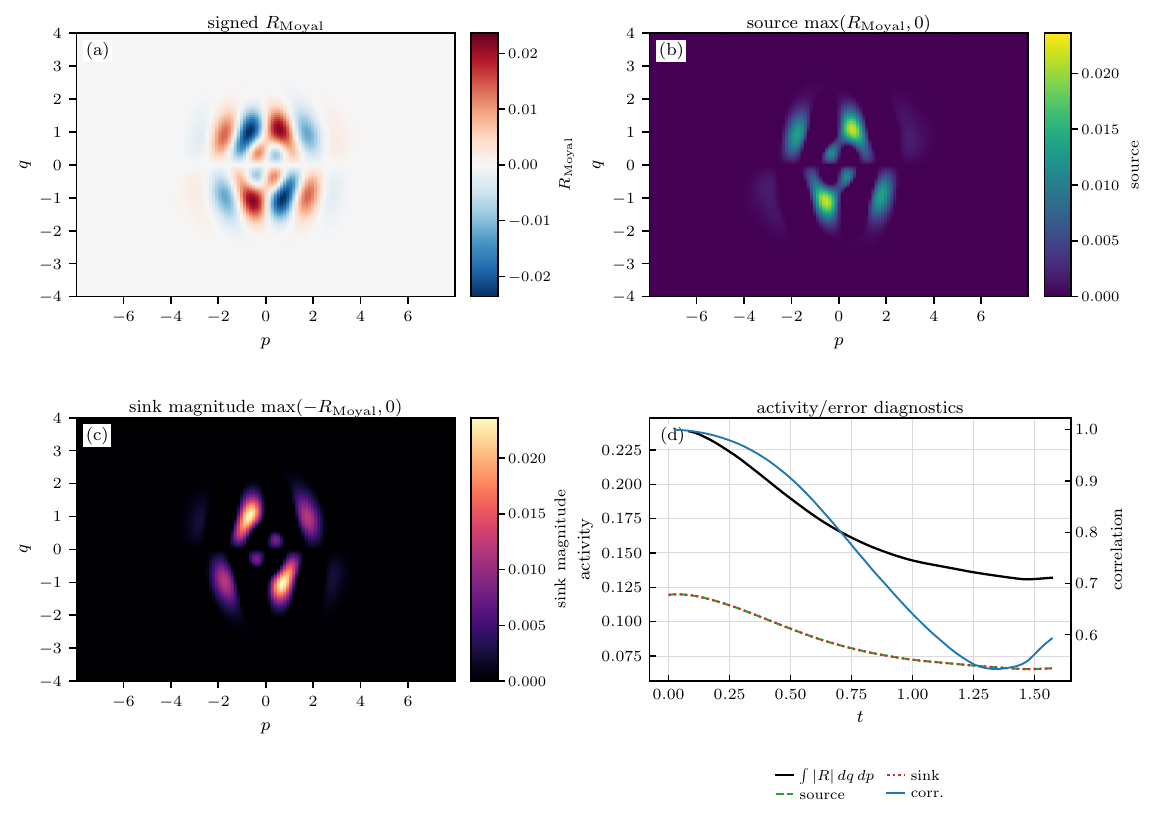}
\caption{Signed Moyal residual for the quartic benchmark. For $V(q)=q^2/2+\lambda q^4$, the nonclassical correction is $R_{\rm Moyal}=-\lambda q\,\partial_p^3 W$ in the dimensionless units used here. The positive and negative parts form a signed source--sink structure in phase space, showing why the correction is not an ordinary positive diffusion process on $(q,p)$.}
\label{fig:quartic_moyal_residual}
\end{figure*}

Figure~\ref{fig:quartic_moyal_residual} isolates the missing generator term itself. Panel (a) shows that $R_{\rm Moyal}$ is a signed source--sink pattern, with alternating positive and negative regions rather than a one-signed rate. Panels (b) and (c) separate the Hahn--Jordan positive and negative parts of the same residual. They show explicitly what a signed-particle or signed-weight representation must sample: not a positive diffusion current, but paired source and sink contributions whose difference reconstructs the third-derivative Moyal term. Panel (d) tracks the total residual activity and the source--sink diagnostics over the evolution, making the signed activity an observed numerical object rather than only a formal operator statement.

\subsection{Measuring the residual strength}
\label{sec:residual_strength_diagnostic}

The same split also gives a compact diagnostic for the size of the term omitted by classical carrier transport. A convenient bounded form is
\[
\chiQ(t)
=
\frac{\|\Qh[W(t)]\|_2}
{\|\Lcl W(t)\|_2+\|\Qh[W(t)]\|_2},
\qquad 0\le \chiQ\le 1,
\]
with the convention $\chiQ=0$ when both norms vanish. When $\|\Lcl W(t)\|_2$ is nonzero, one may equivalently quote the unbounded ratio
\[
\epsilonQ(t)=
\frac{\|\Qh[W(t)]\|_2}{\|\Lcl W(t)\|_2}.
\]
These quantities are not fitted error estimates and not thermodynamic entropy productions. They are generator diagnostics. They vanish identically for quadratic Hamiltonians and become nonzero when the Wigner--Moyal generator contains a signed residual beyond the classical Liouville flow.

For the quartic benchmark, the numerator is precisely the norm of the residual field shown in Fig.~\ref{fig:quartic_moyal_residual}. Thus the diagnostic requires no additional physical approximation: it is computed from the same Moyal residual that the weighted representation must sample. It should also be distinguished from the ordinary truncated-Wigner approximation error. A small time-integrated residual suggests that classical carrier transport may be a controlled approximation for selected observables, but no universal error bound is claimed here without separate stability estimates for the specific dynamics and norm.

\subsection{Reconstructing the signed residual}

The final comparison applies the same classical carrier update and then restores the signed residual. This reconstruction demonstrates the generator split; it is not a positive stochastic process and not a fitted Langevin correction. No smoothing, clipping, damping, positive diffusion coefficient, or stochastic forcing is introduced. For the balanced $\lambda=0.02$ benchmark, the signed residual reduces the final Wigner error from the classical-carrier scale to the numerical residual scale. A larger-$\lambda$ optional stress test at $\lambda=0.05$ gives the same ordering.

\begin{figure*}[t]
\includegraphics[width=\textwidth]{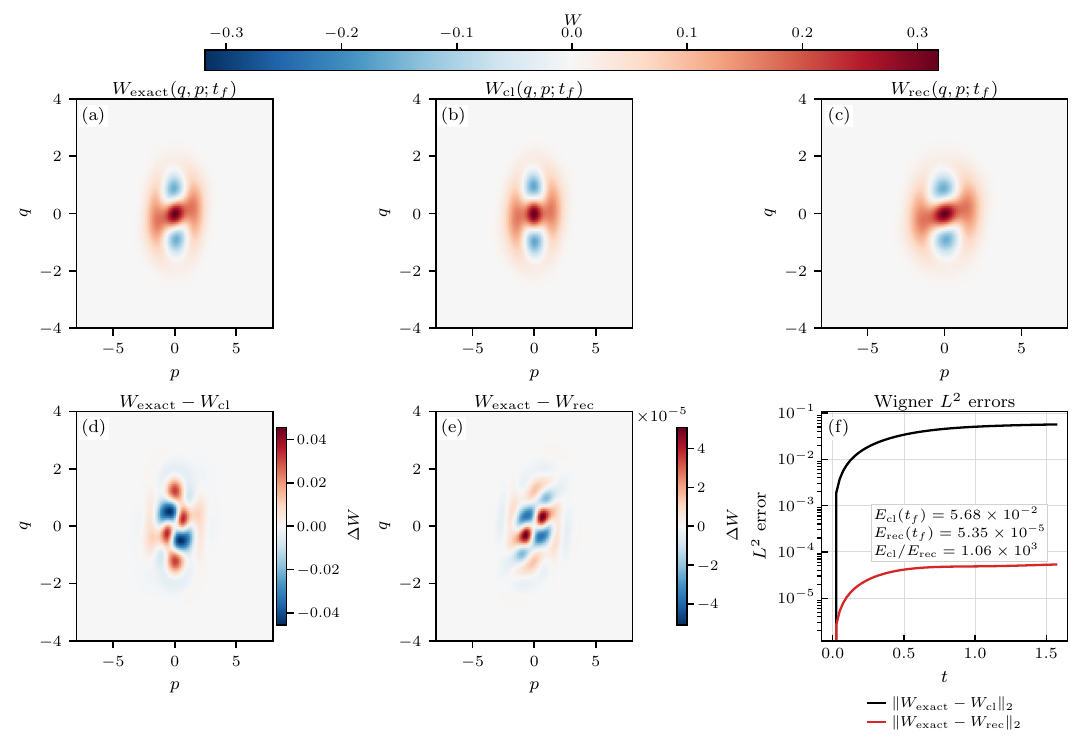}
\caption{Signed-residual reconstruction of the quartic Wigner evolution. Classical carrier transport alone gives a final Wigner error of order $10^{-2}$, while the classical-plus-signed-residual update reduces the error to order $10^{-5}$ for the balanced benchmark. No damping, clipping, smoothing, diffusion coefficient, or fitted Langevin noise is introduced.}
\label{fig:quartic_signed_reconstruction}
\end{figure*}

Figure~\ref{fig:quartic_signed_reconstruction} shows the resulting reconstruction. Panels (a)--(c) compare the exact, classical-only, and signed-residual results at the same final time. The classical-only error in panel (d) retains the same structured source--sink signature seen in the residual plot. After the signed residual is restored, panel (e) shows only a small numerical remainder. The logarithmic error plot in panel (f) gives the scale separation: the classical-only final error is about $5.7\times10^{-2}$, while the residual-corrected error is about $5.4\times10^{-5}$. The improvement is therefore roughly three orders of magnitude. This is the numerical counterpart of the formal statement
\[
L_W^\dagger=L_{\rm cl}^\dagger+\Qh:
\]
the classical carrier gives the reference flow, and the signed residual supplies the missing Wigner--Moyal correction.

\subsection{Numerical checks and tolerances}
\label{sec:numerical_diagnostics}

The balanced grid used $N_q=384$, $N_y=N_p=768$, $dq=dy=0.0625$, $dp\simeq0.131$, $q_{\max}=12$, and $p_{\max}\simeq50.3$. The final time was $t_f=\pi/2$, with time step $dt=0.005$ and 65 saved samples. Normalization and marginal errors remained small: the largest recorded Wigner normalization error was about $1.9\times10^{-11}$, while the wave-function normalization and marginal errors were at or below $3.1\times10^{-14}$. In the harmonic null test the maximum rotation error was $4.8\times10^{-5}$, and the residual norm was zero to machine precision. For the $\lambda=0.02$ quartic benchmark, the exact-versus-classical Wigner $L^2$ error was $5.7\times10^{-2}$, while the signed-reconstruction final error was $5.4\times10^{-5}$, an improvement factor of about $1.1\times10^3$. In the optional $\lambda=0.05$ stress test, the signed error remained about $6.3\times10^{-5}$, compared with a classical-only error of about $1.1\times10^{-1}$. The worst boundary leakage recorded for the balanced manuscript benchmark was below $1.0\times10^{-13}$.

\begin{figure*}[t]
\includegraphics[width=\textwidth]{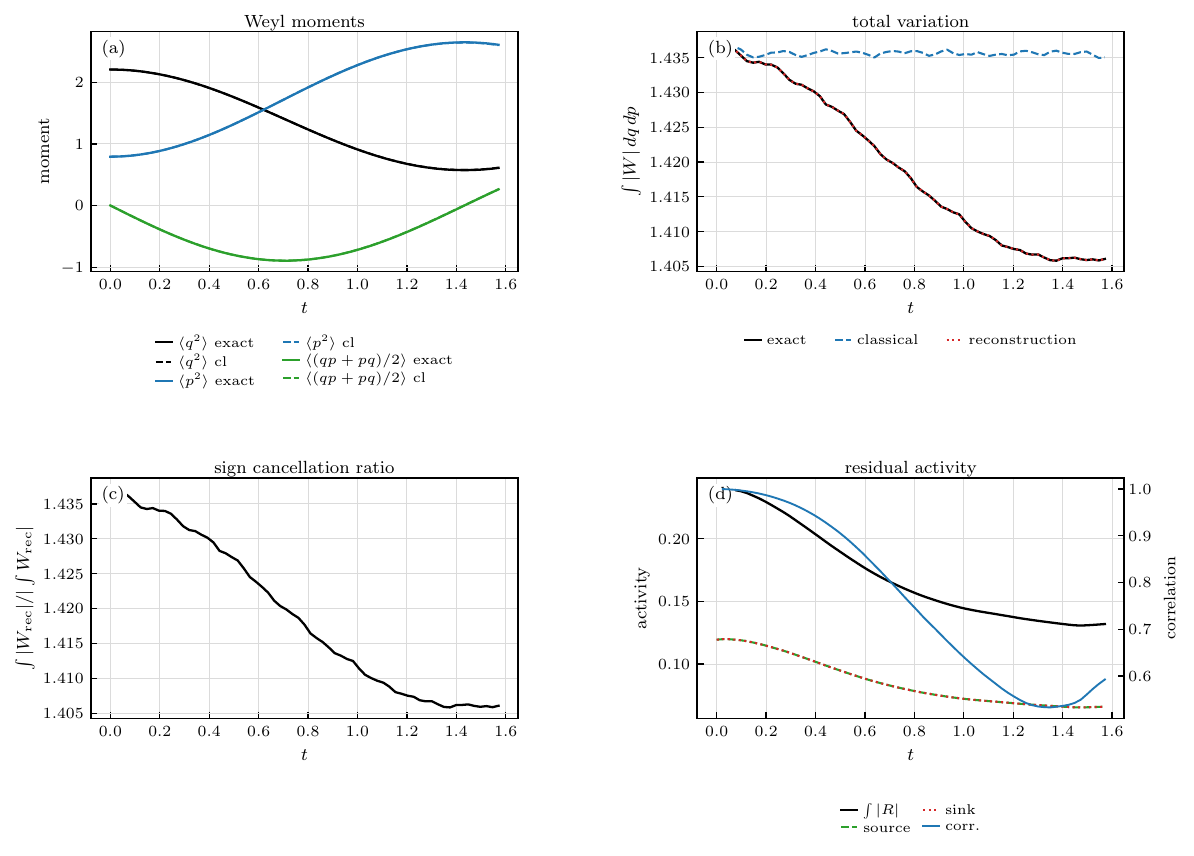}
\caption{Additional diagnostics for the quartic benchmark, including Weyl moments, total variation, sign-cancellation ratio, and residual activity. These diagnostics check that the improvement of the signed-residual reconstruction is not produced by normalization drift, boundary leakage, or artificial damping.}
\label{fig:quartic_numerical_diagnostics}
\end{figure*}

Figure~\ref{fig:quartic_numerical_diagnostics} collects the corresponding moment, total-variation, sign-cancellation, and residual-activity diagnostics. Panel (a) compares selected Weyl moments and checks that the plotted evolution is not being diagnosed only through a final $L^2$ norm. Panels (b) and (c) monitor total variation and sign cancellation, which are the quantities most sensitive to artificial damping, clipping, or hidden loss of signed weight. Panel (d) repeats the residual-activity accounting used in Fig.~\ref{fig:quartic_moyal_residual}; this is the same signed activity entering the residual diagnostic $\chiQ(t)$, up to the choice of norm and normalization. Together with the small normalization and boundary errors quoted above, these diagnostics support the interpretation that the reconstruction improves the Wigner evolution by restoring the signed Moyal residual, not by renormalizing, smoothing, or diffusing the distribution.

\section{Relation to earlier work}

The construction used here touches several established literatures. We now clarify which ingredients are standard and what the present paper adds. The main contribution is the connection
\[
\begin{array}{@{}c@{}}
\text{positive Langevin limitation}
\longrightarrow
\text{Wigner function}
\\[2pt]
\longrightarrow
\text{weighted carrier paths}
\\[2pt]
\longrightarrow
\text{signed path relation}.
\end{array}
\]
This route explains why the free choices in positive coupled phase-space Langevin models are not merely technical defects. They indicate that the target object has been specified too weakly: exact phase-space quantum mechanics is represented by a weighted Wigner measure, not by an unweighted positive trajectory density.

\subsection{Wigner--Moyal mechanics}

The Wigner function and Moyal bracket are standard components of phase-space quantum mechanics \cite{Wigner1932QuantumCorrection,Moyal1949QuantumMechanicsStatisticalTheory,Hillery1984DistributionFunctionsFundamentals}. We do not introduce a new phase-space transform. We use the Wigner function because it is the object selected by the Weyl expectation rule and because its evolution is exactly equivalent to von Neumann dynamics. Reviews of Wigner methods and phase-space quantum dynamics already cover this background in detail \cite{Polkovnikov2010PhaseSpaceRepresentation,SellierNedjalkovDimov2015AppliedWignerMonteCarlo}.

\subsection{Static nonclassicality in phase space}

Wigner negativity and related phase-space criteria are established tools for characterizing nonclassical states \cite{KenfackZyczkowski2004NegativityIndicator,BohmannAgudeloSperling2020PhaseSpaceMatrices}. The present paper uses a different diagnostic distinction. The question is not only whether $\FW(z,t)$ is negative at a given time, but whether the dynamics requires a nonzero signed residual beyond classical carrier transport. This is why the harmonic oscillator is an important null case: the state can be Wigner-negative while the residual generator vanishes. Conversely, an anharmonic Hamiltonian can generate a nonzero signed source--sink residual even when low-dimensional marginals remain close to a classical-carrier approximation.

\subsection{Signed-particle Wigner Monte Carlo}

Signed-particle Wigner Monte Carlo and branching-random-walk methods are also established \cite{Kosina2003MonteCarloCarrierTransport,Nedjalkov2004UnifiedParticleWignerBoltzmann,SellierDimov2014ManyBodyWignerMonteCarlo,Sellier2015SignedParticleFormulation,ShaoXiong2020BranchingRandomWalkWigner,Muscato2021WignerEnsembleNoSplitting,WangSimine2021SignedParticleChemistry}. These methods already show how signed particles or weighted particles can reconstruct Wigner dynamics. In particular, branching treatments based on positive and negative parts of the Wigner kernel are known \cite{ShaoXiong2020BranchingRandomWalkWigner}.

Our use of this machinery is conceptual rather than algorithmic. We begin from the failure of positive phase-space Langevin closure and ask what kind of stochastic representation can replace it. Signed Wigner paths provide one answer: they preserve the ordinary Wigner phase-space representation and exact Wigner--Moyal dynamics by abandoning the requirement that the unweighted empirical density be positive.

\subsection{Nelson stochastic mechanics}

Nelson-type stochastic mechanics provides a positive diffusion process in configuration space \cite{Nelson1966SchrodingerFromNewtonian}. Such processes can reproduce the Born density in $q$-space, but they are not exact positive phase-space representations of quantum dynamics. The known conceptual and mathematical subtleties of Nelson mechanics, including quantization conditions and equivariance issues, are not the focus here \cite{Wallstrom1989StochasticSchrodingerDerivation,Pavon1995HamiltonPrincipleStochasticMechanics,Bacciagaluppi1999NelsonianRevisited}. Here Nelson mechanics serves only as a useful contrast: it keeps positivity by working in configuration space, whereas the present construction keeps the Wigner phase-space target by allowing signed weights.
\[
\begin{array}{@{}c@{\quad}c@{\quad}c@{}}
\begin{array}{c}
\text{positive}\\[-1pt]
+\ \text{configuration space}
\end{array}
&
\Rightarrow
&
\begin{array}{c}
\text{Nelson-type}\\[-1pt]
\text{diffusion}
\end{array}
\\[10pt]
\begin{array}{c}
\text{weighted}\\[-1pt]
+\ \text{Wigner phase space}
\end{array}
&
\Rightarrow
&
\begin{array}{c}
\text{exact Wigner--Moyal}\\[-1pt]
\text{dynamics}
\end{array}.
\end{array}
\]

\subsection{Positive and enlarged phase-space methods}

Another way to avoid Wigner negativity is to change the representation or enlarge the phase space. Positive-$P$ and related methods are important examples \cite{DrummondGardiner1980GeneralisedP,GilchristGardinerDrummond1997PositivePValidity,Deuar2021MultiTimePositiveP}. Modern phase-space stochastic methods also exist for open and many-body systems under appropriate representations or approximations \cite{HuberKirtonRabl2021PhaseSpaceSpinDynamics}. These approaches are not in conflict with the present argument. They illustrate one possible exit: one may keep positivity by changing the represented space or by working with a different phase-space distribution. Our claim concerns the Wigner function itself.

\subsection{Quasiprobability fluctuation relations}

Quantum fluctuation relations and quasiprobability statistics form a large literature. Standard quantum work fluctuation relations already differ from classical ones because work is not represented by a single Hermitian observable in the usual two-time-measurement framework \cite{TalknerLutzHanggi2007WorkNotObservable}. Quasiprobability approaches to dynamical observables, heat, and work show that negative values naturally encode nonclassical features \cite{SolinasGasparinetti2015FullWorkDistribution,SolinasGasparinetti2016InterferenceWorkDistribution,Hofer2017QuasiprobabilityDynamicObservables,Lostaglio2018QuantumFluctuationContextuality,Levy2020QuasiprobabilityHeatFluctuations,Lostaglio2023KirkwoodDiracQuasiprobability,Zhang2024QuasiprobabilityFluctuationTheorem}.

\begin{revtexlayoutpar}
There are also Wigner-space entropy-production fluctuation theorems \cite{Deffner2013QuantumEntropyPhaseSpace}. Our signed Wigner path relation is not a replacement for those thermodynamic results. It is a signed-measure identity associated with the weighted path representation of Wigner--Moyal dynamics. Its distinctive feature is the explicit decomposition
\end{revtexlayoutpar}
\[
\frac{d\mu_F}{d\mu_R^\Theta}
=
\Asign e^{\Amag},
\]
where $\Asign$ is interpreted as an interference-parity factor.

\section{Conclusion}

The attempt to construct a positive coupled Langevin process on quantum phase space is natural, but it does not determine a unique exact Wigner dynamics. Matching the Born position density, or finitely many moments, leaves the drift and diffusion underdetermined. Strengthening the target to all Weyl-ordered expectations fixes the phase-space object as the Wigner function. But the Wigner function is generally signed, and its Moyal evolution is generally not a positive second-order Fokker--Planck flow.

The resolution is to change what the stochastic process is supposed to represent. In the Wigner representation, exact quantum phase-space sampling is not ordinary positive stochasticity. It is weighted stochasticity:
\[
\boxed{
\FW(\z,t)=\E_{\Pp}[W_t\delta(\z-\z_t)].
}
\]
The carrier paths may be classical phase-space trajectories, but the quantum state is their weighted empirical measure. Choosing classical Hamiltonian flow as the carrier isolates the nonclassical contribution in the residual Moyal generator, which vanishes in the formal classical limit.
The numerical benchmarks make this separation explicit: quadratic dynamics transports the signed Wigner function by classical flow, quartic dynamics exposes a finite signed Moyal residual, and restoring that residual recovers the reference Wigner evolution without fitted positive diffusion. The residual diagnostic $\chiQ(t)$ records the same information at the generator level: it is zero for quadratic dynamics and nonzero when classical carrier transport omits a genuine signed Wigner--Moyal source term.

This perspective also changes the meaning of forward--reverse path comparison. For positive path measures, a fluctuation relation compares probabilities. For Wigner paths, the comparison is between signed path contributions:
\[
\frac{d\mu_F}{d\mu_R^\Theta}
=
\Asign e^{\Amag}.
\]
The magnitude term contains carrier and weight asymmetries. The sign factor records the interference parity. It is the part of the comparison with no classical probabilistic analogue. This sign factor also suggests a possible future path-level diagnostic of dynamical nonclassicality, provided the initial Wigner sign is separated from signs generated by residual Moyal activity.

Thus the central lesson is the following three-way tension:
\[
\text{positivity}
+
\text{phase space}
+
\text{exact quantum dynamics}
\]
cannot generally all be maintained. Positive stochasticity can live naturally in configuration space. Exact phase-space quantum dynamics lives naturally in weighted Wigner paths. An exact positive Langevin closure for the Wigner function is not ruled out by preference; it is ruled out by the Wigner--Moyal generator and by the signed nature of generic quantum phase-space states.

\appendix

\section{Technical details}
\label{app:technical_foundations}

\subsection{Why finite moments do not fix a Langevin model}
\label{app:underdetermination}

We give a constructive version of the statement that the marginal does not fix the phase-space process. Let $P(q,p,t)$ be a positive phase-space density satisfying a continuity equation
\[
\partial_t P + \partial_q J_q + \partial_p J_p=0.
\]
For an It\^o diffusion, the current has the form
\[
J_i=A_iP-\frac12\partial_j(D_{ij}P),
\qquad D\ge0,
\]
but the following argument only needs the continuity form.

The position marginal is
\[
\rho(q,t)=\int dp\,P(q,p,t).
\]
Assuming boundary terms vanish as $|p|\to\infty$, integration over $p$ gives
\[
\partial_t\rho(q,t)
+
\partial_q\bar J_q(q,t)=0,
\]
where
\[
\bar J_q(q,t)=\int dp\,J_q(q,p,t).
\]
Thus the marginal dynamics constrains only $\bar J_q$, the $p$-integrated current. It does not determine the full phase-space current $J_q(q,p,t)$.

To make the freedom explicit, let $C(q,p,t)$ be any sufficiently regular function satisfying
\[
\int dp\,\partial_p C(q,p,t)=0
\]
under the same boundary conditions. Then the current transformation
\[
J_q\mapsto J_q+\partial_p C,
\qquad
J_p\mapsto J_p-\partial_q C
\]
leaves the full phase-space continuity equation unchanged because
\[
\partial_q(\partial_p C)+\partial_p(-\partial_q C)=0.
\]
It also leaves the position marginal current unchanged:
\[
\int dp\,(J_q+\partial_p C)
=
\int dp\,J_q.
\]
Thus even if a phase-space current reproduces the correct Born marginal, there is an infinite family of currents that produce the same marginal.

One may also impose finitely many moment constraints. Define
\[
M_k(q,t)=\int dp\,p^kP(q,p,t),
\qquad k=0,1,\dots,N.
\]
The evolution of $M_k$ depends on finitely many weighted integrals of the phase-space currents. Such constraints restrict the allowed currents but do not determine arbitrary functional degrees of freedom in $J_q$ and $J_p$. For example, one may choose current perturbations orthogonal, in the $p$ variable, to the finite span of test functions $\{1,p,\dots,p^N\}$ while maintaining the required boundary behavior. These perturbations leave the specified finite moment equations unchanged.

This establishes the freedom at the level of currents. Translating the current back into drift and diffusion fields is also nonunique. Given a current $J_i$, there are generally many choices of $A_i$ and positive semidefinite $D_{ij}$ satisfying
\[
J_i=A_iP-\frac12\partial_j(D_{ij}P),
\]
provided $P>0$ on the region considered. Therefore finite marginal and moment matching cannot select a unique positive phase-space Langevin process.

This is the mathematical origin of the free functions that appear in constructions based only on reproducing $\rho(q,t)$ and finitely many phase-space moments.

\subsection{Wigner transform identities}
\label{app:wigner_equivalence}

We collect the standard identities used in Section~\ref{sec:wigner_representation}. For a density operator $\hat\rho$, the Wigner transform is
\[
\FW(q,p)
=
\frac{1}{2\pi\hbar}
\int dy\,e^{-ipy/\hbar}
\left\langle q+\frac y2\middle|\hat\rho\middle|q-\frac y2\right\rangle.
\]
Introduce
\[
x=q+\frac y2,
\qquad
x'=q-\frac y2,
\]
so that
\[
q=\frac{x+x'}2,
\qquad
y=x-x'.
\]
Multiplying by $e^{ip(x-x')/\hbar}$ and integrating over $p$ gives
\[
\int dp\,e^{ip(x-x')/\hbar}
\FW\!\left(\frac{x+x'}2,p\right)
=
\left\langle x\middle|\hat\rho\middle|x'\right\rangle,
\]
because
\[
\frac{1}{2\pi\hbar}\int dp\,e^{ip(x-x'-y)/\hbar}
=
\delta(x-x'-y).
\]
Thus the Wigner transform is invertible.

The position marginal follows immediately:
\[
\int dp\,\FW(q,p)
=
\int dy\,\delta(y)
\left\langle q+\frac y2\middle|\hat\rho\middle|q-\frac y2\right\rangle
=
\langle q|\hat\rho|q\rangle.
\]
For a pure state, this is $|\psi(q)|^2$. The momentum marginal is obtained similarly by writing the density matrix in the momentum basis:
\[
\int dq\,\FW(q,p)=\langle p|\hat\rho|p\rangle,
\]
which is $|\tilde\psi(p)|^2$ for a pure state.

The Weyl symbol of an operator $\hat A$ is
\[
A_W(q,p)
=
\int dy\,e^{-ipy/\hbar}
\left\langle q+\frac y2\middle|\hat A\middle|q-\frac y2\right\rangle,
\]
up to the normalization convention paired with the definition of $\FW$. With the convention used here, the trace relation is
\[
\Tr(\hat\rho\hat A)
=
\int dq\,dp\,\FW(q,p)A_W(q,p).
\]
This is the sense in which $\FW$ reproduces all Weyl-ordered expectation values. It is stronger than requiring agreement only of the position and momentum marginals.

For dynamics, start with the von Neumann equation
\[
i\hbar\partial_t\hat\rho=[\hat H,\hat\rho].
\]
Under the Weyl transform, operator products map to the star product:
\[
(\hat A\hat B)_W=A_W\star B_W,
\]
where
\[
\star
=
\operatorname{exp}\!\left[
\frac{i\hbar}{2}
\left(
\overleftarrow{\partial_q}\overrightarrow{\partial_p}
-
\overleftarrow{\partial_p}\overrightarrow{\partial_q}
\right)
\right].
\]
Thus
\[
\partial_t\FW
=
-\frac{i}{\hbar}\left(H\star \FW-\FW\star H\right)
=
\{H,\FW\}_M,
\]
where the Moyal bracket is
\[
\{H,\FW\}_M
=
\frac{1}{i\hbar}(H\star\FW-\FW\star H).
\]
For pure states, the von Neumann equation for $\hat\rho=|\psi\rangle\langle\psi|$ is equivalent to the Schr\"odinger equation for $|\psi\rangle$ up to a time-dependent global phase. This establishes the equivalence chain used in the main text.

\subsection{Moyal equation versus Fokker--Planck}
\label{app:fokker_planck_limit}

For
\[
H(q,p)=\frac{p^2}{2m}+V(q),
\]
the Moyal bracket expansion yields
\[
\begin{aligned}
\partial_t\FW
&=
-\frac{p}{m}\partial_q\FW
+
\sum_{n=0}^{\infty}
\frac{(-1)^n}{(2n+1)!}
\left(\frac{\hbar}{2}\right)^{2n}
\\
&\qquad{}\times
V^{(2n+1)}(q)\partial_p^{2n+1}\FW.
\end{aligned}
\]
The $n=0$ term is
\[
V'(q)\partial_p\FW.
\]
Thus
\[
\partial_t\FW
=
-\frac{p}{m}\partial_q\FW
+
V'(q)\partial_p\FW
+
\Qh[\FW],
\]
where
\[
\Qh[\FW]
=
\sum_{n=1}^{\infty}
\frac{(-1)^n}{(2n+1)!}
\left(\frac{\hbar}{2}\right)^{2n}
V^{(2n+1)}(q)\partial_p^{2n+1}\FW.
\]

An ordinary It\^o diffusion on phase space has a Fokker--Planck equation
\[
\partial_tP
=
-\partial_i(A_iP)
+
\frac12\partial_i\partial_j(D_{ij}P),
\qquad D=BB^T\ge0.
\]
This is a local second-order differential operator in the phase-space variables. It preserves positivity of $P$ under standard regularity and boundary assumptions.

For a potential with $V'''(q)\neq0$, the Wigner--Moyal generator contains the term
\[
-\frac{1}{3!}
\left(\frac{\hbar}{2}\right)^2V'''(q)\partial_p^3\FW.
\]
No choice of drift $A_i$ and positive diffusion matrix $D_{ij}$ can reproduce this third derivative as part of an ordinary Brownian Fokker--Planck generator for arbitrary $\FW$. Higher derivatives appear whenever higher odd derivatives of $V$ are nonzero. Equivalently, the Wigner potential can be written in a nonlocal momentum-transfer form with a signed kernel rather than a positive transition rate.

Therefore, for generic non-quadratic potentials, exact Wigner--Moyal evolution is not an ordinary positive Brownian diffusion on phase space.

This statement has important limitations. It does not rule out:
\begingroup
\hbadness=10000
\begin{enumerate}[label=(\roman*),leftmargin=*]
  \item exact signed or weighted particle representations;
  \item nonlocal jump processes with signed weights;
  \item complex stochastic processes;
  \item enlarged phase-space representations such as positive-$P$ type methods;
  \item approximate methods such as truncated Wigner dynamics;
  \item special quadratic Hamiltonians, for which the Moyal residual vanishes.
\end{enumerate}
\endgroup
The claim is only that the exact Wigner function cannot generally be the probability density of an ordinary positive Brownian phase-space Langevin process.

\subsection{Minimal signed activity for a fixed residual}
\label{app:jordan_activity}

Let $K$ be a signed measure or signed kernel, suppressing variables for notational clarity. When $K$ is absolutely continuous with respect to a positive reference measure, we also write $K$ for its signed density.

A positive decomposition of $K$ is a pair of nonnegative measures or kernels $K_1,K_2\ge0$ such that
\[
K=K_1-K_2.
\]
The Hahn--Jordan decomposition is
\[
K=K^+-K^-,
\]
where $K^+,K^-\ge0$ are mutually singular and satisfy
\[
|K|=K^++K^-.
\]
In the density case this becomes
\[
K^+=\max(K,0),
\qquad
K^- =\max(-K,0).
\]

We claim that this decomposition minimizes total variation activity among positive decompositions. If $K=K_1-K_2$ with $K_1,K_2\ge0$, then the total variation measure satisfies
\[
|K|(A)\le K_1(A)+K_2(A)
\]
for every measurable set $A$. The Hahn--Jordan decomposition attains equality because $|K|=K^++K^-$. In the density case, this reduces pointwise to $K_1+K_2\ge |K|$. After integration over the relevant variables, the total variation activity is therefore minimal.

Here this result is used only after the residual kernel has been fixed. It is not a claim of global uniqueness across all possible representations or all possible carrier processes. Different choices of carrier generator $L_0^\dagger$ lead to different residual kernels. Once the classical carrier and residual kernel are fixed, however, the Hahn--Jordan split is the natural minimal-total-variation decomposition of that residual kernel.

\subsection{Signed path ratio}
\label{app:path_asymmetry}

Let $\Omega$ be a measurable path space. Let $\Pp_F$ and $\Pp_R^\Theta$ be positive probability measures on the same path space. Let $W_F$ and $W_R^\Theta$ be real measurable weights. Define signed measures
\[
d\mu_F=W_F\,d\Pp_F,
\qquad
d\mu_R^\Theta=W_R^\Theta\,d\Pp_R^\Theta.
\]
Here $W_R^\Theta(\GammaPath)$ denotes $W_R(\GammaPath^\dagger)$ after pulling the reversed process back to the forward path space.

Assume the comparison is restricted to a common support $\Omega_0\subset\Omega$ on which
\[
W_F\neq0,
\qquad
W_R^\Theta\neq0,
\]
and where $\Pp_F$ is absolutely continuous with respect to $\Pp_R^\Theta$. Then the Radon--Nikodym derivative of the signed measures exists on this support in the form
\[
\frac{d\mu_F}{d\mu_R^\Theta}
=
\frac{W_F}{W_R^\Theta}
\frac{d\Pp_F}{d\Pp_R^\Theta}.
\]
Write
\[
W_F=\operatorname{sgn}(W_F)|W_F|,
\qquad
W_R^\Theta=\operatorname{sgn}(W_R^\Theta)|W_R^\Theta|.
\]
Then
\[
\frac{W_F}{W_R^\Theta}
=
\operatorname{sgn}(W_F)\operatorname{sgn}(W_R^\Theta)
\frac{|W_F|}{|W_R^\Theta|}.
\]
Define
\[
\Asign
=
\operatorname{sgn}(W_F)\operatorname{sgn}(W_R^\Theta),
\]
and
\[
\Amag
=
\log\frac{d\Pp_F}{d\Pp_R^\Theta}
+
\log\frac{|W_F|}{|W_R^\Theta|}.
\]
Then
\[
\frac{d\mu_F}{d\mu_R^\Theta}
=
\Asign e^{\Amag}.
\]

The corresponding integral identity follows by inversion:
\[
d\mu_R^\Theta
=
\Asign e^{-\Amag}d\mu_F.
\]
For any test functional $\Psi$,
\[
\int_{\Omega_0}\Psi\,d\mu_R^\Theta
=
\int_{\Omega_0}\Psi\,\Asign e^{-\Amag}\,d\mu_F.
\]
Taking $\Psi=1$ gives
\[
\int_{\Omega_0}d\mu_R^\Theta
=
\int_{\Omega_0}\Asign e^{-\Amag}\,d\mu_F.
\]
If the singular parts outside $\Omega_0$ vanish and the two signed measures have the same total normalization, then
\[
\frac{\int_{\Omega_0}\Asign e^{-\Amag}\,d\mu_F}
{\int_{\Omega_0}d\mu_F}
=
1.
\]

Equivalently, in terms of the positive forward sampling measure,
\[
\int_{\Omega_0}
W_F\Asign e^{-\Amag}
\,d\Pp_F
=
\int_{\Omega_0}
W_R^\Theta\,d\Pp_R^\Theta.
\]
This is often the safest form because the expectation is taken with respect to the positive carrier measure $\Pp_F$.

Zero-weight paths and singular components require separate treatment. This is analogous to the absolute-continuity caveats in ordinary fluctuation relations, but with the additional complication that the measures here are signed. In the main text, the relation is therefore stated on the common nonzero-weight support.

\section{Relation to fluctuation theorems and implementation}
\label{app:boundary_implementation}

\subsection{How this differs from fluctuation theorems}
\label{app:fluctuation_literature}

\begin{revtexlayoutpar}
We spell out the relation between the signed Wigner path identity and existing fluctuation-theorem literature. The purpose is not to review quantum thermodynamics comprehensively, but to delimit the claim made here.
\end{revtexlayoutpar}

\subsubsection{Classical fluctuation relations}

Classical fluctuation relations compare positive probability measures over forward and reversed histories. In their path-ratio form, the relevant object is typically
\[
\log\frac{d\Pp_F}{d\Pp_R^\Theta},
\]
which may be identified with total entropy production under appropriate thermodynamic assumptions. The key point is that both measures are positive. The logarithm is therefore a real path functional, and the corresponding integral identities are expectations with respect to ordinary probability measures.

The signed Wigner relation differs at this first step. The object being compared is not only the positive carrier measure $\Pp_F$, but the signed measure
\[
d\mu_F=W_Fd\Pp_F.
\]
Consequently the forward--reverse comparison contains both a magnitude term and a sign term:
\[
\frac{d\mu_F}{d\mu_R^\Theta}
=
\Asign e^{\Amag}.
\]
The sign factor $\Asign$ has no classical counterpart.

\subsubsection{Quantum work relations}

Quantum work fluctuation theorems are often formulated using the two-point measurement scheme. A central lesson of that literature is that work is not represented by a single ordinary Hermitian observable in the same way as energy \cite{TalknerLutzHanggi2007WorkNotObservable}. Coherent initial states and incompatible measurement contexts naturally lead to modified statistics or quasiprobability descriptions \cite{SolinasGasparinetti2015FullWorkDistribution,SolinasGasparinetti2016InterferenceWorkDistribution}.

The present work does not define work, heat, or dissipated energy. It also does not introduce projective energy measurements. The path relation is instead a comparison between signed Wigner path measures associated with phase-space representations of unitary dynamics. Thus it should not be read as a replacement for Crooks or Jarzynski relations in quantum thermodynamics.

\subsubsection{Wigner-space entropy production}

Wigner phase space has already been used to formulate entropy-production fluctuation theorems in open or thermodynamic settings \cite{Deffner2013QuantumEntropyPhaseSpace}. Such work is directly relevant because it shows that Wigner methods can support fluctuation-theorem structures. However, the object considered here is different.

Here the primary structure is
\[
\FW(z,T)
=
\E_{\Pp}\!\left[W[\GammaPath]\delta(z-z_T)\right],
\]
and the path relation compares the signed measures generated by this representation. No heat bath, stationary nonequilibrium state, dissipative generator, or thermodynamic entropy functional is assumed. The sign factor is interpreted as the interference parity, not entropy production.

\subsubsection{Quasiprobability fluctuation frameworks}

\begin{revtexlayoutpar}
A broader quantum-information and quantum-thermodynamics literature uses quasiprobabilities to describe fluctuations of noncommuting observables and coherent protocols \cite{Hofer2017QuasiprobabilityDynamicObservables,Lostaglio2018QuantumFluctuationContextuality,Levy2020QuasiprobabilityHeatFluctuations,Lostaglio2023KirkwoodDiracQuasiprobability,Zhang2024QuasiprobabilityFluctuationTheorem}. This literature is important here because it establishes that negative or signed fluctuation weights are not merely pathologies. They may encode contextuality, incompatibility, interference, or other nonclassical features.
\end{revtexlayoutpar}

Our relation is consistent with that viewpoint, but it is more specific. It is tied to a Wigner carrier-path representation and decomposes the forward--reverse signed-measure ratio into
\[
\Amag
=
\log\frac{d\Pp_F}{d\Pp_R^\Theta}
+
\log\frac{|W_F|}{|W_R^\Theta|}
\]
and
\[
\Asign
=
\operatorname{sgn}(W_F)\operatorname{sgn}(W_R^\Theta).
\]
The first part contains the positive carrier-law asymmetry and the absolute weight asymmetry. The second records whether the forward and reversed path contributions have equal or opposite Wigner sign.

\subsubsection{Signed-particle Wigner methods}

Signed-particle Wigner Monte Carlo and branching-random-walk methods already represent Wigner dynamics through particles or paths carrying positive and negative signs \cite{Sellier2015SignedParticleFormulation,ShaoXiong2020BranchingRandomWalkWigner,Muscato2021WignerEnsembleNoSplitting,WangSimine2021SignedParticleChemistry}. These methods are algorithmic and probabilistic antecedents of the carrier-path picture used here. In particular, positive/negative kernel decompositions of the Wigner potential operator are known.

The new element here is not signed-particle dynamics itself, but the forward--reverse signed-measure comparison applied to such weighted Wigner carrier paths, together with the explicit magnitude/sign decomposition and the interpretation of the sign as the interference parity.

\subsubsection{What the identity does and does not say}

The signed Wigner path relation should be understood as a signed-measure identity:
\[
\frac{d\mu_F}{d\mu_R^\Theta}
=
\Asign e^{\Amag}.
\]
It becomes fluctuation-theorem-like when integrated:
\[
\int d\mu_R^\Theta
=
\int \Asign e^{-\Amag}d\mu_F.
\]
But unless additional thermodynamic structure is supplied, this is not an entropy-production theorem.

The claim is therefore limited and precise:
\begin{quote}
\begin{revtexlayoutpar}
For weighted Wigner carrier paths, forward--reverse comparison naturally decomposes into a positive magnitude asymmetry and an interference-parity factor. The sign factor records the part of the path-reversal accounting carried by Wigner interference rather than by positive probability.
\end{revtexlayoutpar}
\end{quote}

\subsection{Numerical implementation notes}
\label{app:numerics}

A direct numerical implementation of the representation separates three tasks. First, the carrier dynamics is generated by the classical Hamiltonian flow. Second, the residual Wigner potential operator is represented either by a signed momentum-transfer kernel, by a signed grid update, or by an equivalent branching construction. Third, observables are reconstructed from weighted contributions rather than from the unweighted empirical density.

For the benchmarks discussed in the main text, the harmonic oscillator is used as a null test. Since $\Qh=0$, the evolution reduces to deterministic classical rotation of the initial Wigner function, including any initial negativity. The quartic oscillator tests the first nontrivial residual term, proportional to $-\hbar^2\lambda q\,\partial_p^3\FW$, and therefore probes the signed part of the dynamics.

The grid benchmarks reported here use marginal errors, selected Weyl-moment errors, total variation, sign-cancellation ratio, residual activity, and the residual diagnostic $\chiQ(t)$ defined in Sec.~\ref{sec:residual_strength_diagnostic}. In path-sampling implementations, the empirical distributions of $\Amag$ and $\Asign$ would provide further diagnostics. These diagnostics distinguish three failures that can otherwise be conflated: sampling variance, sign cancellation, and replacement of the exact Wigner generator by a positive approximate closure.



\hbadness=10000
\bibliography{references_weighted_wigner_paths_enhanced}

\end{document}